\magnification 1200
\centerline {\bf  Can the Quantum Measurement Problem be resolved }
\vskip 0.3cm
\centerline {\bf within the framework of Schroedinger Dynamics and Quantum 
Probability?}
\vskip 1cm
\centerline {\bf by Geoffrey Sewell}   
\vskip 0.5cm
\centerline {\it Department of Physics, Queen Mary, University of London}
\vskip 0.1cm
\centerline {\it Mile End Road, London E1 4NS, UK. e-mail: g.l.sewell@qmul.ac.uk}
\vskip 1cm\noindent
{\bf Abstract} We provide an affirmative answer to the question posed in the title. Our 
argument is based on a treatment of the Schroedinger dynamics of the composite, 
$S_{c}$, of a quantum microsystem, $S$, and a macroscopic measuring apparatus, 
${\cal I}$, consisting of $N$ interacting particles. The pointer positions of this apparatus 
are represented by subspaces of its representative Hilbert space that are simultaneous 
eigenspaces of coarse-grained intercommuting macroscopic observables. By taking 
explicit account of their macroscopicality, we prove that, for suitably designed apparatus 
${\cal I}$, the evolution of the composite $S_{c}$ leads both to the reduction of the 
wave-packet of $S$ and to a one-to-one correspondence between the resultant state of 
this microsystem and the pointer position of ${\cal I}$, up to utterly neglible corrections 
that decrease exponentially with $N$. 
\vskip 1cm
\centerline {\bf 1. Introductory Discussion } 
\vskip 0.3cm
According toVon Neumann\rq s [1] phenomenological picture of the measurement 
process, the coupling of a microsystem, $S$, to a measuring instrument, ${\cal I}$,  leads 
to the following two effects. 
\vskip 0.1cm\noindent
(I) It converts a pure state of $S$, as given by a linear combination 
${\sum}_{r=1}^{n}c_{r}u_{r}$ of its orthonormal energy eigenstates $u_{r}$, into a 
statistical mixture of these states for which ${\vert}c_{r}{\vert}^{2}$ is the probability 
of finding this system in the state $u_{r}$. This effect is often termed the \lq reduction of 
the wave packet\rq . 
\vskip 0.1cm\noindent
(II) It sends a certain set of classical, i.e. intercommuting, macroscopic variables $M$ of  
${\cal I}$ to values, indicated by pointers,  that specify which of the states $u_{r}$ of 
$S$ is realised. 
\vskip 0.1cm\noindent
This leaves us with the following basic question of quantum measurement theory, which 
is clearly pertinent to that of the completeness of quantum mechanics.
\vskip 0.1cm\noindent
{\it ${\cal Q}$. Can the standard quantum dynamics of the composite $(S+{\cal I})$, 
allied to a suitable choice of the macroscopic observables $M$, lead to the effects (I) and 
(II)?} 
\vskip 0.1cm
This is the contentious question that we address here and, as in [2,3], our response to it is 
affirmative, as is that of some other authors [4-6]. Our present objective is to describe the 
mathematical structure of our general argument that leads to this conclusion and that is 
supported by treatments  of some concrete models. We remark that our affirmative 
response to ${\cal Q}$ conflicts with that of Von Neumann [1] and Wigner [7], who 
argued that the observation of  the pointer of ${\cal I}$ requires another measuring 
instrument,  ${\cal I}_{2}$, which in turn requires yet another instrument, and so on, in 
such a way that the whole process involves an infinite regression ending up in the 
observer\rq s brain!\footnote*{Others have taken the view that the measurement problem 
cannot be resolved without modification of the Schroedinger dynamics of $S_{c}$,  due 
to {\it either} its interaction with the \lq rest of the Universe\rq\  [8-11] 
{\it or} a certain postulated nonlinearity that leads to a classical deterministic evolution of 
its macroscopic observables [12].} In our view, the essential flaw in the Von Neumann-
Wigner theory is its failure to take explicit account of the macroscopicality of the 
observables corresponding to the pointer positions of  the measuring apparatus. By 
contrast, we build this macroscopicality into our treatment, and thereby evince a 
qualitative difference between the characteristic properties of the microsystem $S$ and 
the macrosystem ${\cal I}$ that removes the need for a Von Neumann-Wigner 
regression. 
\vskip 0.1cm
As a preliminary to setting up a mathematical theory of the measurement process, we 
note that it is clear from the works of Bohr [13], Jauch [14] and Van Kampen [5] that 
such a theory demands both a characterisation of the macroscopicality of the observables 
$M$ and an amplification property of the $S-{\cal I}$ coupling whereby different 
microstates of $S$ give rise to macroscopically different states of ${\cal I}$. Evidently, 
this implies that the initial state in which ${\cal I}$ is prepared must be unstable against 
microscopic changes in the state of $S$. On the other hand, as emphasised by Whitten-
Wolfe and Emch [15, 16],  the correspondence between the microstate of $S$ and the 
eventual observed macrostate of ${\cal I}$ must be stable against macroscopically small 
changes in the initial state of this instrument, of the kind that are inevitable in 
experimental procedures. Thus, the system $(S+{\cal I})$ must exhibit a striking 
combination of stability and instability properties.
\vskip 0.1cm   
Rigorous constructive treatments of the measurement process, which take account of the 
above considerations, have been provided by Hepp [17] and Whitten-Wolfe and Emch 
[15, 16] on the basis of models for which  ${\cal I}$ is idealised as an infinitely extended 
system of finite density, for which the macroscopic observables $M$ are intercommuting 
global intensive ones. This picture of ${\cal I}$ corresponds to that employed for the 
statistical mechanical description of large systems in the thermodynamic limit [18-20], 
and it has the merit of sharply distinguishing between macroscopically different states, 
since different values of $M$ correspond to disjoint primary representations of the 
observables. Moreover, in the treatments of the measurement problem based on this 
picture, the models of Hepp [17] and Whitten-Wolfe and Emch [15, 16] do indeed exhibit 
the required reduction of the wave-packet and the one-to-one correspondence between 
the pointer position of ${\cal I}$  and the resultant state of $S$; and these results are 
stable against all localised perturbations of the initial state of ${\cal I}$. On the debit 
side, however, Hepp\rq s model requires an infinite time for the measurement to be 
effected (cf. Bell [21]), while although that of Whitten-Wolfe and Emch achieves its 
measurements in finite times, it does so only by dint of a physically unnatural, globally 
extended $S-{\cal I}$ interaction.
\vskip 0.1cm
In view of these observations, we base our treatment of the measuring process on the  
generic model of $S_{c}=(S+{\cal I})$ for which ${\cal I}$ is a large but finite $N$-
particle system. Our main aim is to determine whether there are viable conditions under 
which this model yields the same essential results as the infinite one, but within a finite 
realistic observational time.  In fact, we have achieved this aim in recent works [2, 3],  in 
which we showed that the quantum mechanics of the finite model does indeed contain the 
structures required for an affirmative answer to the question ${\cal Q}$, as illustrated by 
an explicit treatment of the finite version of the Coleman-Hepp model [17]. This result 
provides rigorous mathematical substantiation of the arguments of Refs. [4-6] which led 
to the same main conclusion.  A key feature of our treatment is the representation  of the 
macro-observables $M$ and the pointer positions of  ${\cal I}$ within the framework 
proposed by Van Kampen [22] and Emch [23], whereby $M$ comprises a set of coarse-
grained intercommuting extensive observables and  the pointer positions of ${\cal I}$ 
correspond to their simultaneous eigenspaces. These are mutually orthogonal subspaces 
of the Hilbert space of the pure states of ${\cal I}$ and are the natural analogues of 
classical phase cells. Most importantly, their dimensionalities are astronomically large, 
increasing exponentially with $N$, and their role in the present treatment is analogous to 
that of the inequivalent representation spaces in the infinite system models of Refs. [15-
17]. As a result, the finite system model yields the essential positive results of the infinite 
one, but within a finite, realistic observational time. To be precise, it exhibits the above 
properties (I) and (II), in a stable manner, up to corrections that decrease exponentially 
with $N$ and that are therefore utterly negligible by any standards of experimental 
physics. 
\vskip 0.1cm
The present note is devoted to a description of our mathematical scheme that has led to 
these results. We start in Section 2 by constructing the generic model of $S_{c}$ and 
formulating both the time-dependent expectation values of the observables of $S$ and 
their conditional expectation values, given the values of the 
macro-observables $M$ of ${\cal I}$, subject to the assumption that $S$ and ${\cal I}$ 
are independently prepared and then coupled together at time $t=0$. In Section 3, we 
formulate the conditions on the dynamics of the model and the structure of the macro-
observables $M$ under which it exhibits the properties (I) and (II) of a measurement 
process in a stable manner. We then briefly describe two models that fulfill these 
conditions, referring the reader to other articles for their detailed treatments. In Section 4, 
we probe further into the probabilistic structure of the model in order to pin-point, in 
general terms, the source of the effects (I) and (II).  There we find that this is provided by 
a {\it large deviation} principle, which  represents a rather general collective property of 
many-particle systems [24] and constitutes a generalisation to certain nonequilibrium  
states of Einstein\rq s formula, $P={\rm const.}{\rm exp}(S)$, which relates the 
equilibrium probability distribution, $P$, of  chosen macroscopic variables to the entropy 
function, $S$, of these variables. We conclude, in Section 5, with a brief  resume of the 
picture presented here.
\vskip 0.5cm
\centerline {\bf 2. The Generic Model }
\vskip 0.3cm
We assume that the algebras of bounded observables of the microsystem $S$, the 
instrument  ${\cal I}$ and their composite $S_{c}=(S+{\cal I})$ are those of the 
bounded operators in separable Hilbert spaces  ${\cal H}, {\cal K}$ and ${\cal 
H}{\otimes}{\cal K}$, respectively. Correspondingly, the states of  these systems are 
represented by the density matrices in the respective spaces. The density matrices for the 
pure states of any of these systems are then the projection operators $P(f)$ of their 
normalised vectors $f$. For simplicity we assume that ${\cal H}$ is of finite 
dimensionality $n$.
\vskip 0.1cm
We base the macroscopic description of ${\cal I}$ pertinent to the measuring process on 
an abelian subalgebra ${\cal M}$ of ${\cal B}$, which is generated by coarse-grained 
macroscopic observables (cf. [22, 23]): these are typically extensive variables of parts or 
the whole of ${\cal I}$. The choice of ${\cal M}$ yields a partition of ${\cal K}$ into the 
simultaneous eigenspaces ${\cal K}_{\alpha}$ of its elements. These subspaces of ${\cal 
K}$ are termed  \lq phase cells\rq\ as they are the canonical analogues of classical phase 
cells. We take them to represent the macrostates of ${\cal I}$ and we assume that they 
are unequivocally indicated by the \lq pointer positions\rq\ of this instrument. Most 
importantly, the dimensionality of each ${\cal K}_{\alpha}$ is astronomically large, 
since it increases exponentially with the corresponding entropy and thus with $N$. 
\vskip 0.1cm
Since ${\cal I}$ is designed so that the pointer readings are in one-to-one correspondence 
with the eigenstates $u_{1},. \ .,u_{n}$ of $S$, we assume that the index ${\alpha}$ of 
its macrostates also runs from $1$ to $n$. Hence, denoting the projection operator for 
${\cal K}_{\alpha}$ by ${\Pi}_{\alpha}$, it follows from the above specifications that 
$${\Pi}_{\alpha}{\Pi}_{\beta}={\Pi}_{\alpha}{\delta}_{{\alpha}{\beta}},\eqno(1)$$ 
$${\sum}_{{\alpha}=1}^{n}{\Pi}_{\alpha}=I_{\cal K}\eqno(2)$$
and that each element $M$ of ${\cal M}$ takes the form 
$$M={\sum}_{{\alpha}=1}^{n}M_{\alpha}{\Pi}_{\alpha},\eqno(3)$$
where the $M_{\alpha}$\rq s are scalars. 
\vskip 0.1cm
We assume that $S_{c}$ is a conservative system, whose Hamiltonian operator $H_{c}$, 
in ${\cal H}{\otimes}{\cal K}$, takes the form
$$H_{c}=H{\otimes}I_{\cal K}+I_{\cal H}{\otimes}K+V,\eqno(4)$$
where $H$ and $K$ are the Hamiltonians of $S$ and ${\cal I}$, respectively, and  $V$ is 
the $S-{\cal I}$ interaction energy. Further, we assume that ${\cal I}$ is an instrument of 
the first kind [14], in that the interaction $V$ induces no transitions between the 
eigenstates of $H$. Thus, since the latter comprise an orthogonal basis set $(u_{1},. \ 
.,u_{n})$ of ${\cal H}$, with energy levels $({\epsilon}_{1},. \ .,{\epsilon}_{n})$, 
respectively, the operators $H$ and $V$ take the forms 
${\sum}_{r=1}^{n}{\epsilon}_{r}P(u_{r})$ and 
${\sum}_{r=1}^{n}P(u_{r}){\otimes}V_{r}$, respectively, where the $V_{r}$\rq s are 
self-adjoint operators in ${\cal K}$. Hence, by Eq. (4),  $H_{c}$ reduces to the form 
$H_{c}={\sum}_{r=1}^{n}P(u_{r}){\otimes}K_{r},$ where
$$K_{r}=K+V_{r}+{\epsilon}_{r}I_{\cal K}.\eqno(5)$$
Consequently the one-parameter unitary group 
${\lbrace}U_{c}(t)={\rm exp}(iH_{c}t){\vert}t{\in}{\bf R}{\rbrace}$, which governs 
the dynamics of $S_{c}$, is given by the formula 
$$U_{c}(t)={\sum}_{r=1}^{n}P(u_{r}){\otimes}U_{r}(t),
\eqno(6)$$
where
$$U_{r}(t)={\rm exp}(iK_{r}t).\eqno(7)$$
\vskip 0.3cm
{\bf Note.} Eqs (5)-(7) signify that different eigenstates $u_{r}$ of $S$ give rise to 
different evolutions of ${\cal I}$. This is crucial to the realisability of the effect (II) of 
Sec. 1, whereby the pointer of ${\cal I}$ is driven into a position determined by the 
eigenstate of $S$.
\vskip 0.3cm
{\bf Initial Conditions.} We assume that the the systems $S$ and ${\cal I}$ are coupled 
together at time $t=0$ following independent preparation of $S$ in a pure state and 
${\cal I}$ in a mixed one, as represented by a normalised vector ${\psi}$ and a density 
matrix ${\Omega}$, respectively. Thus the initial state of the composite $S_{c}$ is given 
by the density matrix 
$${\Phi}=P({\psi}){\otimes}{\Omega}\eqno(8)$$
and its evolute at time $t \ (>0)$ is 
$$U_{c}^{\star}(t){\Phi}U_{c}(t):={\Phi}(t).\eqno(9)$$
Further, since ${\psi}$ is a normalised vector, it is a linear combination of the basis 
vectors $(u_{1},. \  .,u_{n})$ and hence takes the form 
$${\psi}={\sum}_{r=1}^{n}c_{r}u_{r},\eqno(10)$$
where ${\sum}_{r=1}^{n}{\vert}c_{r}{\vert}^{2}=1$. Hence, by Eqs. (8)-(10),
$${\Phi}(t)={\sum}_{r,s=1}^{n}{\overline c}_{r}c_{s}P_{r,s}
{\otimes}{\Omega}_{r,s}(t),\eqno(11)$$
where $P_{r,s}$ is the operator in ${\cal H}$ defined by the equation
$$P_{r,s}f=(u_{s},f)u_{r} \ {\forall} \ f{\in}{\cal H}\eqno(12)$$
and  
$${\Omega}_{r,s}(t)=U_{r}^{\star}(t){\Omega}U_{s}(t).\eqno(13)$$
We note that ${\Omega}_{r,r}(t)$ is just the evolute of ${\Omega}$ corresponding to 
the eigenstate $u_{r}$ of $S$.
\vskip 0.3cm
{\bf Expectation and Conditional Expectation Values of Observables.} The observables 
of $S_{c}$ with which we shall be concerned are just the self-adjoint elements of ${\cal 
A}{\otimes}{\cal M}$. Their expectation values for the time-dependent state ${\Phi}(t)$ 
are given by the formula
$$E\bigl(A{\otimes}M\bigr)={\rm Tr}\bigl({\Phi}(t)[A{\otimes}M]\bigr) \ {\forall} \ 
A{\in}{\cal A}, \ M{\in}{\cal M}.\eqno(14)$$
In particular, the expectation values of the observables of $S$ are given by the equation
$$E(A)=E(A{\otimes}I_{\cal K})  \ {\forall} A{\in}{\cal A},\eqno(15)$$
while the probability that the macrostate of ${\cal I}$ corresponds to the cell ${\cal 
K}_{\alpha}$ is 
$$w_{\alpha}=E(I_{\cal H}{\otimes}{\Pi}_{\alpha}).\eqno(16)$$
Further, in view of the abelian property of the algebra ${\cal M}$, the functional $E$ 
induces a conditional expectation of $A({\in}{\cal A})$ with respect to ${\cal M}$, 
namely a linear map, $E(.{\vert}{\cal M})$, of ${\cal A}$ into ${\cal M}$ that preserves 
positivity and normalisation and that satisfies the compatibility condition 
$$E\bigl(E(A{\vert}{\cal M})M\bigr)=E(A{\otimes}M) \ {\forall} \ A{\in}{\cal A}, \ 
M{\in}{\cal M}.\eqno(17)$$
Moreover, since $E(A{\vert}{\cal M})$ is an element of ${\cal M}$ and therefore takes 
the form ${\sum}_{{\alpha}=1}^{n}f_{\alpha}(A){\Pi}_{\alpha}$, where the 
$f_{\alpha}$\rq s are linear, positive, normalised functionals on ${\cal A}$, it follows 
from this observation and Eqs. (1)-(3), (16) and (17) that 
$f_{\alpha}(A)=E(A{\otimes}{\Pi}_{\alpha})/w_{\alpha}$ and hence that
$$E(A{\vert}{\cal M})={\sum}_{{\alpha}=1}^{n}
E(A{\otimes}{\Pi}_{\alpha}){\Pi}_{\alpha}/w_{\alpha}.$$
Since ${\Pi}_{\alpha}$ is the projector for the cell ${\cal K}_{\alpha}$, its coefficient is 
the conditional expectation value $E(A{\vert}{\cal  K}_{\alpha})$ of $A$, given the 
macrostate ${\cal K}_{\alpha}$ of ${\cal I}$. Thus
 $$E(A{\vert}{\cal K}_{\alpha})=E(A{\otimes}{\Pi}_{\alpha})/w_{\alpha} \ 
{\forall} \ A{\in}{\cal A}, \ w_{\alpha}{\neq}0.\eqno(18)$$
\vskip 0.1cm
We now relate $E(A), \ E(A{\vert}{\cal K}_{\alpha})$ and $w_{\alpha}$ to the key 
dynamical quantity
$$F_{r,s:{\alpha}}=Tr\bigl({\Omega}_{r,s}(t){\Pi}_{\alpha}\bigr) \ 
{\forall} \ r,s,{\alpha}{\in}{\lbrace}1,. \ .,n{\rbrace},\eqno(19)$$
noting in particular that $F_{r,r;{\alpha}}$ is the probability that ${\cal K}_{\alpha}$ is 
the macrostate of ${\cal I}$ at time $t$ when $u_{r}$ is the state of $S$. 
By Eqs. (2), (11)-(13), (16), (18) and (19), $E(A), \ E(A{\vert}{\cal K}_{\alpha})$ and 
$w_{\alpha}$ are related to $F$ by the formulae 
$$E(A)={\sum}_{r=1}^{n}{\vert}c_{r}{\vert}^{2}(u_{r},Au_{r})+
{\sum}_{r{\neq}s;r,s=1}^{n}{\sum}_{{\alpha}=1}^{n}F_{r,s:{\alpha}}
{\overline c}_{r}c_{s}(u_{r},Au_{s}) \ {\forall} \ A{\in}{\cal A},\eqno(20)$$
$$E(A{\vert}{\cal K}_{\alpha})={\sum}_{r,s=1}^{n} F_{r,s;{\alpha}}
{\overline c}_{r}c_{s}(u_{r},Au_{s}) /w_{\alpha}\ {\forall} \ A{\in}
{\cal A}, \ w_{\alpha}{\neq}0.\eqno(21)$$
and 
$$w_{\alpha}={\sum}_{r,s=1}^{n}{\overline c}_{r}c_{s}F_{r,s;{\alpha}}.
\eqno(22)$$
The following key properties of $F$ ensue from Eqns. (2), (13) and (19).  
$${\sum}_{{\alpha}=1}^{n}F_{r,r:{\alpha}}=1,\eqno(23)$$
$$1{\geq}F_{r,r:{\alpha}}{\geq}0,\eqno(24)$$
$$F_{r,s:{\alpha}}={\overline F}_{s,r:{\alpha}}\eqno(25)$$
and, for $z_{1},. \ .,z_{n}{\in}{\bf C}$, the sesquilinear form ${\sum}_{r,s=1}^{n}
{\overline z}_{r}z_{s}F_{r,s;{\alpha}}$ is positive. Hence
$$F_{r,r;{\alpha}}F_{s,s;{\alpha}}{\geq}{\vert}F_{r,s;{\alpha}}{\vert}^{2}.
\eqno(26)$$
\vskip 0.5cm
\centerline {\bf 3. The Measurement Process }
\vskip 0.3cm
Suppose now that a reading of the pointer of ${\cal I}$ is made at time $t$. Then, 
according to the standard probabilistic interpretation of quantum mechanics, it follows 
from the above specifications that
\vskip 0.1cm\noindent
(i) $E(A)$ is the expectation value of  the observable $A$ of $S$ immediately before the 
reading;
\vskip 0.1cm\noindent
(ii) $w_{\alpha}$ is the probability that the reading yields the result that the macrostate 
of ${\cal I}$ corresponds to the cell ${\cal K}_{\alpha}$; and,
\vskip 0.1cm\noindent
(iii) in that case,  $E(A{\vert}{\cal K}_{\alpha})$ is the expectation value of $A$ 
immediately after the measurement.
\vskip 0.1cm\noindent
Hence the conditions for the realisation of the demands (I) and (II) of Von Neumann\rq s 
phenomenological picture, described in Sec. 1, are that, for $t$ greater than some critical, 
realistic value, ${\tau}$, and less, in order of magnitude, than the Poincare' recurrence 
times, 
\vskip 0.1cm\noindent
(a) $$E(A)={\sum}_{r=1}^{n}{\vert}c_{r}{\vert}^{2}(u_{r},Au_{r}) \ {\forall} \ 
A{\in}{\cal A} \eqno(27)$$
and 
\vskip 0.1cm\noindent
(b) there is a unique invertible transformation ${\phi}$ of the set $(1,2,. \ .,n)$ such that
$$E(A{\vert}{\cal K}_{\alpha})=(u_{{\phi}({\alpha})},Au_{{\phi}({\alpha})}) \ 
{\forall} \ A{\in}{\cal A}. \eqno(28)$$
In other words, the pointer reading ${\alpha}$ signifies that the resultant state of $S$ is 
$u_{{\phi}({\alpha})}$. 
\vskip 0.3cm
{\bf Proposition.} {\it The combination of conditions (a) and (b) is equivalent to the 
formula
$$F_{r,r;{\phi}^{-1}(r)}=1 \ {\forall} \ r=1,. \ .,n.\eqno(29)$$
Hence this formula is equivalent to the Von Neumann conditions (I) and (II).}
\vskip 0.3cm
{\bf Proof.} By Eqs. (24) and (26), Eq. (29) implies that $F_{r,s;{\alpha}}$ vanishes 
unless $r=s={\phi}({\alpha})$ and is therefore equivalent to the formula
$$F_{r,s;{\alpha}}={\delta}_{r,{\phi}({\alpha})}{\delta}_{s,{\phi}({\alpha})} \ 
{\forall} \ r,s,{\alpha}=1,. \ .,n.\eqno(30)$$
Assuming this formula , it follows immediately from Eqs. (20)-(22) that conditions (a) 
and (b) are fulfilled. Conversely, assuming condition (b), it follows from a comparison of 
the form of  $w_{\alpha}E(A{\vert}{\cal K}_{\alpha})$  obtained from Eqs. (22) and 
(28) with that given by Eq. (21) that
$${\sum}_{r,s=1}^{n}{\overline c}_{r}c_{s}F_{r,s;{\alpha}}
(u_{{\phi}({\alpha}},Au_{{\phi}({\alpha})})={\sum}_{r,s=1}^{n}
{\overline c}_{r}c_{s}F_{r,s;{\alpha}}(u_{r},Au_{s}) \ {\forall} \ A{\in}{\cal A}, \ 
c_{1}, \  ,c_{n}{\in}{\bf C}.$$
On equating coefficients of ${\overline c}_{r}c_{s}$ in this formula, we see that
$$F_{r,s;{\alpha}}\bigl[(u_{{\phi}({\alpha})},Au_{{\phi}({\alpha})})-
(u_{r},Au_{s})\bigr]=0 \ {\forall} \ A{\in}{\cal A}, \ r,s=1,2.. \ ,n.$$
In view of the orthonormality of the $u_{r}$\rq s and the invertibility of ${\phi}$, this 
last formula is equivalent to Eq. (30) and thus to Eq. (29). 
\vskip 0.3cm
{\bf Weakening of the Condition (29).} In fact, that condition is extremely stringent and 
one sees both from the study of tractable models [2-4] and from a general probabilistic 
argument, presented in Sec. 4, that it should be weakened by a corrective term, due to 
macroscopic fluctuations, that decreases exponentially with $N$ for large $N$. In other 
words, the sharp condition given by Eq. (29) should be weakened to the form wherein the 
difference between the two sides of that equation does not exceed ${\rm exp}(-cN)$, 
where $c$ a positive constant of the order of unity. Hence, in view of Eqs. (23) and (24), 
this weaker condition is that
$$0{\leq}1-F_{r,r;{\phi}^{-1}(r)}{\leq}{\rm exp}(-cN).\eqno(31)$$
Correspondingly, it follows [2, 3] from arguments parallel to those employed to pass 
from Eqs. (27) and (28) to Eq. (30) in the proof of the above Proposition that the 
weakening of  condition (29) to Eq. (31) implies corrections of order ${\rm exp}(-cN/2)$ 
to Eqs. (27) and (28), i.e. to the Von Neumann conditions (I) and (II). These corrections 
are utterly negligible from a physical standpoint, since $N$ is typically of the order of 
$10^{8d}$, where $d$ is the dimensionality of the instrument ${\cal I}$.
\vskip 0.1cm
A further essential property of an efficacious measuring instrument is that it should be 
stable against local perturbations of its initial state ${\Omega}$ (cf. [15, 16]). We express 
this condition in the following form.
\vskip 0.1cm\noindent 
${\cal S}$. {\it  The formula (31) is stable under perturbations of the initial state 
${\Omega}$ of ${\cal I}$ that are localised in the sense of leaving this state unchanged 
outside a ball whose volume is $O(1)$ with respect to $N$.}
\vskip 0.1cm
Thus, we characterise a quantum measurement process by the conditions given by Eq. 
(31) and ${\cal S}$. These conditions are viable, since they have been shown to be 
fulfilled by concrete models that may be described as follows: we refer the reader to the 
cited articles for full treatments of them. 
\vskip 0.3cm
{\bf Model 1} [2, 3]. This is a finite version of the Coleman-Hepp model [17], which is 
designed to measure the $z$-component, say, of the spin of an electron. The model  
consists of an electron, ${\cal E}$, and a chain, ${\cal C}$, of  $N$ Pauli spins. We 
regard the electron ${\cal E}$ as a composite of its own Pauli spin, $S$, and a spinless 
particle, ${\cal P}$, that carries its orbital motion. We then take the spin $S$ to be the 
microsystem under observation and the composite of ${\cal P}$ and ${\cal C}$ to be the 
measuring instrument ${\cal I}$. We take the phase cells of ${\cal I}$ to be the 
subspaces ${\cal K}_{\pm}$ of its canonically defined Hilbert space ${\cal K}$ that 
correspond to positive and negative polarisations, respectively. We assume that ${\cal 
E}$ and ${\cal C}$ are coupled together at $t=0$ following independent preparations of 
$S$ and ${\cal P}$ in pure states and ${\cal C}$ in a state of equilibrium\footnote*{This 
is the state that maximises the entropy of ${\cal C}$ subject to the specified constraint, 
and by the subadditivity of entropy [25] it takes the form ${\otimes}_{n=1}^{N}{1\over 
2}(I+m{\sigma}_{n,z})$, where $m$ is its polarisation and ${\sigma}_{n,x}$ is the $z$-
component of the $n$\rq th spin of ${\cal C}$.} subject to a constraint, subsequently 
removed, that fixes its polarisation to a value directed along $Oz$: the initial state of  
${\cal P}$ is assumed to be a wave packet localised at one end of ${\cal C}$ and moving 
towards the other end of that chain. The  ${\cal E}-{\cal C}$ coupling, following this 
preparation, is assumed to be of the form $P_{-}{\otimes}V$, where $P_{-}$ is the 
projection operator for the eigenstate of $S$ for which the $z$-component, $s_{z}$, of 
its spin is negative and $V$ is an interaction between ${\cal P}$ and the spins comprising 
${\cal C}$ that reverses the latter ones in turn as the electron passes by them. Thus, up to 
corrections due to fluctuation effects, the resultant macrostate of ${\cal I}$ corresponds 
to the cell ${\cal K}_{+}$ or ${\cal K}_{-}$ according to whether $s_{z}$ is in its 
eigenstate with positive or negative eigenvalue. To be precise [2, 3], taking fluctuations 
properly into account, the system ${\cal I}=({\cal P}+{\cal C})$ satisfies the conditions 
given by Eq. (31) and ${\cal S}$ and is therefore an effective measuring instrument for 
the spin $S$; while the time taken to effect the measurement is essentially that required 
for the electron to run the length of the chain.
\vskip 0.3cm
{\bf Model 2}.  This is the Allahverdyan-Balian-Nieuwenhuizen (ABN) model [4]. It 
consists of a localised Pauli spin, $S$, an Ising-Weiss ferromagnet, ${\cal F}$, and a heat 
bath ${\cal B}$. As described in the terms of the present article, $S$ is the microsystem 
under observation and the composite $({\cal F}+{\cal B})$ is the instrument ${\cal I}$ 
that measures the $z$-component, $s_{z}$, of its spin. The phase cells of ${\cal I}$ are 
the subspaces ${\cal K}_{\pm}$ of its canonically defined Hilbert space ${\cal K}$ that 
carry positive and negative polarisations, respectively, along $Oz$. It is assumed that the 
initial states of $S, \ {\cal F}$ and ${\cal B}$ are uncorrelated with ${\cal F}$ in a 
polarisation-free metastable state below its transition temperature and ${\cal B}$ in a 
thermal equilibrium state at that temperature. It is assumed that the subsequent couplings 
of the components of the model comprise (a) long range interactions between the $z$-
components of the spin of $S$ and those of  ${\cal F}$ and (b) interactions between 
${\cal B}$ and ${\cal F}$ that drive the latter into an equilibrium state with positive or 
negative polarisation along $Oz$ according to whether $s_{z}$ is positive or negative. 
Under these conditions, ${\cal I}$ serves as an effective measuring apparatus for $S$ [4], 
i.e. it satisfies Eq. (31) and the stability condition\footnote*{This condition was not 
treated in [4], but a simple application of the method of Refs. [2, 3] demonstrates that the 
ABN model does satisfy it}. Again, as in Model 1, the cells ${\cal K}_{\pm}$ represent 
the pointer positions for the eigenstates of the $s_{z}$  with positive and negative 
eigenvalues, respectively.
\vskip 0.5cm
\centerline {\bf  4. Role of the Large Deviation Principle}
 \vskip 0.3cm
We now aim to probe more deeply into the quantum statistics of the measurement process 
in order to excavate the source of the properties of ${\cal I}$ represented by Eq. (31) and 
the stability condition ${\cal S}$. For simplicity, we confine our attention here to the 
case where the pointer reading corresponds to the value of a single coarse grained 
macroscopic observable\footnote{**}{Generalisation to the case of several 
intercommuting ones is straightforward.} $M$ of ${\cal I}$ and thus where the algebra 
${\cal M}$ is generated by bounded functions of $M$.  Further, we assume that $M$ is a 
coarsened version of a fine-grained extensive observable $M_{f}$ in the following sense 
(cf. [22, 23]).  
\vskip 0.1cm
Defining the intensive variable
$$m:=N^{-1}M_{f},\eqno(32)$$
we assume that $m$ is a bounded operator in ${\cal K}$ with pure point spectrum, 
whose highest and lowest points are the extremals of a closed interval ${\Delta}$ of ${\bf 
R}$. Further, we assume that this spectrum simulates a continuum for large $N$, in that 
the maximum spacing between the eigenvalues of $m$ tends to zero as $N$ tends to 
infinity. 
\vskip 0.1cm
We construct the coarse-grained version $M$ of $M_{f}$ by dividing ${\Delta}$ into a 
set of disjoint intervals of equal length ${\lbrace}{\cal J}_{\alpha}{\vert}{\alpha}=1, \  
.,n{\rbrace}$ and defining ${\cal K}_{\alpha}$ to be the subspace of ${\cal K}$ spanned 
by the eigenvectors of $m$ whose eigenvalues lie in ${\cal J}_{\alpha}$. We then define 
$m_{\alpha}$ to be the arithmetic mean of those eigenvalues and define the coarse-
grained observable $M$ by the formula
$$M={\sum}_{{\alpha}=1}^{n}Nm_{\alpha}{\Pi}_{\alpha},\eqno(33)$$
where, as previously, ${\Pi}_{\alpha}$ is the projection operator for ${\cal 
K}_{\alpha}$.
\vskip 0.1cm
We recall that, as noted following Eq. (19), $F_{r,r;{\alpha}}$ is the probability that 
${\cal K}_{\alpha}$ is the macrostate of ${\cal I}$ at time $t$ that corresponds to the 
eigenstate $u_{r}$ of $S$.  Furthermore, the density matrix ${\Omega}_{r,r}(t)$, 
defined by Eq. (13), represents the state of the {\it full system} ${\cal I}$ at time $t$ 
corresponding to this eigenstate of $S$. Hence it follows from the above construction of 
$M$ in terms of the fine-grained intensive observable $m$, which we may treat as a 
classical variable, that 
$$F_{r,r;{\alpha}}=P_{r}(m{\in}{\cal J}_{\alpha}),\eqno(34)$$
where $P_{r}$ is the probability measure on ${\Delta}$ induced by the state 
${\Omega}_{r,r}(t)$. 
\vskip 0.1cm
We now assume that this probability satisfies a {\it large deviation principle} [24], which 
is widely applicable to intensive macroscopic variables of many-particle systems. As 
noted in Sec. 1, this is a natural generalisation to nonequilibrium situations of Einstein\rq 
s formula $P={\rm const.}{\rm exp}(S)$ for the relationship at thermal equilibrium 
between  the entropy $S$, expressed as function of macroscopic variables and their 
probability distribution $P$. Thus, in the present situation, the large deviation principle 
asserts that the density of the probability measure $P_{r}$ on the variable $m$ takes the 
form 
$f_{r,N}(m){\rm exp}\bigl(N{\sigma}_{r}(m)\bigr)$, where the function 
${\sigma}_{r}$ is $N$-independent and plays the role of a generalised specific entropy 
and $N^{-1}{\rm log}\bigl(f_{N,r}(m)\bigr)$ tends to zero as $N$ tends to infinity. 
Thus, for large $N$, the density of  $P_{r}$ is effectively governed by the exponential 
term ${\rm exp}(N{\sigma}_{r})$ . 
\vskip 0.1cm
In order to demonstrate that, under very mild conditions on ${\sigma}_{r}$, both Eq. 
(31) and the stability condition ${\cal S}$ are satisfied, we suppose that
\vskip 0.1cm\noindent
(a) for each $r{\in}(1,2, \ .,n), \ {\sigma}_{r}$ attains its maximum at precisely one 
value, $m_{r}$, of $m$;
\vskip 0.1cm\noindent
(b) $m_{r}$ lies in the interior of one of the intervals ${\cal J}_{\alpha}$, namely ${\cal 
J}_{{\phi}^{-1}(r)}$, and ${\phi}^{-1}(r){\neq}{\phi}^{-1}(s)$ if $r{\neq}s$; 
\vskip 0.1cm\noindent
(c) ${\sigma}_{r}(m_{r})-{\sigma}_{r}(m)$ is greater than some positive constant $c$  
for all $r{\in}(1,2, \ .,n)$ and $m{\notin}{\cal J}_{{\phi}^{-1}(r)}$; and
\vskip 0.1cm\noindent
(d) the function ${\sigma}_{r}$ is unaffected by localised perturbations, as specified in 
condition ${\cal S}$, of the initial state ${\Omega}$ of ${\cal I}$.
\vskip 0.1cm\noindent
These conditions are certainly viable and, indeed, they are fulfilled by the models of 
Refs. [2-4]. Further, when taken in conjunction with the above specified large deviation 
principle for $P_{r}$, they imply that $a$ is an invertible transformation of the set $(1,2, 
\ .,n)$ and that 
$$P_{r}(m{\in}{\cal J}_{\alpha})<{\rm exp}(-cN) \ {\rm for} \ {\alpha}{\neq}
{\phi}^{-1}(r),$$
where $c$ is a positive constant of the order of unity. In view of Eq. (34), this signifies 
that the instrument ${\cal I}$ satisfies the condition (3); and property (d) ensures that it 
fulfills the stability condition ${\cal S}$.
\vskip 0.5cm
\centerline {\bf  5. Concluding Remarks}
\vskip 0.3cm
We have provided a general mathematical scheme for the description of quantum 
measurement theory within the framework of Schroedinger dynamics and quantum 
probability. This scheme is manifestly realisable and involves nothing more than the 
dynamics of the composite of a quantum microsystem $S$ and a macroscopic measuring 
instrument ${\cal I}$. Thus, it does not require any appeal either to cosmological actions 
or to Von Neumann-Wigner psycho-physical parallelism, with its infinite hierarchy of 
measuring instruments. Indeed the only participation of human intelligence in the 
measurement process is in the design of the instrument ${\cal I}$ and the interpretation  
of readings of its pointer positions.  
\vskip 0.5cm
\centerline {\bf References} 
\vskip 0.3cm\noindent
[1] J. Von Neumann: {\it Mathematical Foundations of Quantum Mechanics}, Princeton 
University Press, Princeton, NJ, 195
\vskip 0.1cm\noindent
[2] G. L.Sewell: Rep. Math. Phys. {\bf 56}, 271, 2005
\vskip 0.1cm\noindent
[3] G. L. Sewell: Markov Processes and Rel. Fields {\bf 13}, 425, 2007
\vskip 0.1cm\noindent
[4] A. E. Allahverdyan, R. Balian and Th. M. Nieuwenhuizen. Eur. Phys. Lett. {\bf 61}, 
452, 2003
\vskip 0.1cm\noindent 
[5] N. G. Van Kampen: {\it Physica} A {\bf 153}, 97, 1988.
\vskip 0.1cm\noindent
[6] A. Peres: Am. J. Phys. {\bf 54}, 688 , 1986
\vskip 0.1cm\noindent
[7] E. P.Wigner: Pp. 171-84 of {\it Symmetries and Reflections}, Indiana University 
Press, Bloomington, 1967.
\vskip 0.1cm\noindent
[8]  N. Gisin: {\it Phys. Rev. Lett.} {\bf 52}, 1657, 1984
\vskip 0.1cm\noindent
[9] E. Joos and H. D. Zeh: {\it Z. Phys.} B {\bf 59}, 223,  1985
\vskip 0.1cm\noindent
[10] L. Diosi: {\it J. Phys.} A {\bf 21}, 2885, 1988
\vskip 0.1cm\noindent
[11] I. Percival: {\it Quantum State Diffusion}, Cambridge Univ. Press,  Cambridge, 
1998.
\vskip 0.1cm\noindent
[12] G. C. Ghirardi, A. Rimini and T. Weber: {\it Phys. Rev.} D {\bf 34}, 470, 1986
\vskip 0.1cm\noindent
[13]  N. Bohr: {\it Discussion with Einstein on epistomological problems in atomic 
physics}, Pp. 200-241 of  {\it Albert Einstein: Philosopher-Scientist}, Ed. P. A. Schilp, 
The Library of Living Philosophers, Evanston, IL, 1949.
\vskip 0.1cm\noindent
[14] J, M. Jauch: {\it Foundations of Quantum Mechanics}, Addison Wesley, Reading, 
MA, 1968.
\vskip 0.1cm\noindent
[15] B. Whitten-Wolfe and G. G. Emch: {\it Helv. Phys. Acta} {\bf 49}, 45, 1976
\vskip 0.1cm\noindent
[16] G. G. Emch: Pp. 255-264 of  {\it Quantum Information and Communication},  E. 
Donkor, A. R. Pirich and H. E. Brandt, Eds., Intern. Soc. Opt. Eng. (SPIE) Proceedings 
5105, 2003
\vskip 0.1cm\noindent
[17] K.Hepp: {\it Helv. Phys. Acta} {\bf 45}, 237, 1972
\vskip 0.1cm\noindent
[18] D. Ruelle: {\it Statistical Mechanics}, W. A. Benjamin, New York, 1969.
\vskip 0.1cm\noindent
[19] G. G. Emch: {\it Algebraic methods in Statistical Mechanics and Quantum Field 
Theory},  Wiley, New York, 1972.
\vskip 0.1cm\noindent
[20] G. L. Sewell: {\it Quantum Mechanics and its Emergent Macrophysics}, Princeton 
University Press, Princeton, 2002.
\vskip 0.1cm\noindent
[21]  J. S. Bell: {\it Helv. Phys. Acta} {\bf 48}, 93, 1975
\vskip 0.1cm\noindent
[22] N. G. Van Kampen: {\it Physica} {\bf 20}, 603, 1954
\vskip 0.1cm\noindent
[23] G. G. Emch: {\it Helv. Phys. Acta} {\bf 37}, 532, 1964
\vskip 0.1cm\noindent
[24] R. S. Ellis: {\it Entropy, Large Deviations and Statistical Mechanics}, Springer, New 
York, 1985
\vskip 0.1cm\noindent
[25] E. H. Lieb and M. B. Ruskai: J. Math. Phys. {\bf 14}, 1938, 1973
\end